\documentclass[aps,pra,twocolumn, superscriptaddress, showpacs]{revtex4-1}
\usepackage{hyperref}
\usepackage{amsmath}
\usepackage{amssymb,amsfonts}
\usepackage{graphicx}
\usepackage{relsize}
\usepackage{lmodern}
\usepackage{scalefnt}
\usepackage{subfigure}
\usepackage{xspace} 
\usepackage{ifpdf}
\usepackage{pgffor}
\usepackage{array} 
\usepackage{stackrel}
\newcolumntype{C}{>{$}c<{$}} 
\usepackage{tikz}
\usetikzlibrary{shapes.geometric}
\usepackage{color}
\usetikzlibrary{arrows,matrix,calc,scopes,decorations.markings}
\allowdisplaybreaks[1]

\makeatletter
\newcommand*{\Relbarfill@}{\arrowfill@\Relbar\Relbar\Relbar}
\newcommand*{\xeq}[2][]{\ext@arrow 0055\Relbarfill@{#1}{#2}}
\makeatother
\newcommand{\be}{ \begin{equation}}
\newcommand{\ee}{\end{equation}}

\begin{document}

\title{Emergent Einstein Equation in $p$-adic CFT Tensor Networks}
\author{Lin Chen}
\author{Xirong Liu}
\affiliation{State Key Laboratory of Surface Physics, Fudan University, 200433 Shanghai, China}
\affiliation{Shanghai Qi Zhi Institute, 41st Floor, AI Tower, No. 701 Yunjin Road, Xuhui District, Shanghai, 200232, China}
\affiliation{Department of Physics and Center for Field Theory and Particle Physics, Fudan University, Shanghai 200433, China}
\author{Ling-Yan Hung}
\email{lyhung@fudan.edu.cn}
\affiliation{State Key Laboratory of Surface Physics, Fudan University, 200433 Shanghai, China}
\affiliation{Shanghai Qi Zhi Institute, 41st Floor, AI Tower, No. 701 Yunjin Road, Xuhui District, Shanghai, 200232, China}
\affiliation{Department of Physics and Center for Field Theory and Particle Physics, Fudan University, Shanghai 200433, China}
\affiliation{Institute for Nanoelectronic devices and Quantum computing, Fudan University, 200433 Shanghai, China}
\date{\today}
    

\begin{abstract}
We take the tensor network describing explicit $p$-adic CFT partition functions proposed in \cite{Hung:2019zsk}, and considered boundary conditions of the network describing a deformed Bruhat-Tits (BT) tree geometry. We demonstrate that this geometry satisfies an emergent graph Einstein equation in a {\it unique} way that is consistent with the bulk effective matter action encoding the same correlation function as the tensor network, at least in the perturbative limit away from the {\it pure} BT tree.  Moreover, the (perturbative) definition of the graph curvature in the Mathematics literature \cite{Yau1,Ollivier,Gubser:2016htz} naturally emerges from the consistency requirements of the emergent Einstein equation. This could provide new insights into the understanding of gravitational dynamics potentially encoded in more general tensor networks. 

\end{abstract}
\pacs{11.15.-q, 71.10.-w, 05.30.Pr, 71.10.Hf, 02.10.Kn, 02.20.Uw}
\maketitle

The AdS/CFT \cite{Maldacena:1997re} provided deep insights of (quantum) gravity. 
One extremely important breakthrough inspired by the Ryu-Takayanagi entanglement formula \cite{Ryu:2006bv} is the realization that  (semi-classical) geometries as solutions of the gravity theory are basically geometrization of the entanglement structure of wave-functions of the dual CFT. (see for example \cite{VanRaamsdonk:2016exw} for a review and references therein.)  Tensor networks (TN) widely used to construct many-body wave-functions are also geometrization of patterns of entanglement of many-body wave-functions.  This led to proposals that TNs similar to the MERA captures the microscopic mechanism behind the AdS/CFT correspondence \cite{Swingle:2009bg}. Toy models have been constructed \cite{Pastawski:2015qua, Hayden:2016cfa}  that recreate many aspects of the AdS/CFT correspondence, most notably the RT formula and the error correcting property \cite{Almheiri:2014lwa}.  TN also provides deep insight in the surface-state correspondence \cite{Miyaji:2015fia}, kinematic space \cite{Czech:2015kbp},
 the complexity of the wave-functions and their evolution \cite{Susskind:2014rva, Brown:2015lvg, Caputa:2017urj,Caputa:2017yrh,Czech:2017ryf} and the island formula that is probably the key to the black hole information paradox \cite{Penington:2019npb, Almheiri:2019hni}. 

Nonetheless, reconstruction of gravitational dynamics using TN remains a tremendous challenge. There is some progress by considering relative entropies\cite{Faulkner:2013ica,Faulkner:2014jva} and also complexity optimisation\cite{Caputa:2017urj,Caputa:2017yrh,Czech:2017ryf}, although it is not completely clear how bulk matter or generic time dependence can be included (some progress is recently made in \cite{Boruch:2020wax}.).  

In this paper, we present the emergence of a {\it graph} Einstein equation based on the proposed TN \cite{Hung:2019zsk} of the p-adic AdS/CFT  \cite{Gubser:2016guj,Heydeman:2016ldy}.  Our strategy is to parametrise the form of edge distances, curvatures and stress tensor using an ansatz that is based only on locality and symmetry. We will show that {\it if} an Einstein equation exists at all the self-consistency constraints are stringent enough to return a {\it unique} assignment of bulk geometry from the TN. 
This is perhaps the first such quantitative demonstration involving both matter and time \footnote{Here time corresponds to an extra coordinate.  p-adic CFT is basically a Euclidean theory with no distinction of space and time.},  and where both the CFT and the TN can be explicitly defined, taking advantage of the simplicity of $p$-adic CFT. 

{\bf \begin{center} Lightning review of $p$-adic CFTs \end{center}} 
A one dimensional $p$-adic CFT lives in the $p$-adic number field $Q_{p}$ for any given prime number $p$. i.e. coordinates $x \in Q_{p}$. 
$p$-adic numbers $Q_p$ are field extensions of the rational numbers alternative to the reals $\mathbb{R}$.  This can be readily generalized to an $n$-dimensional $p$-adic CFT by considering field extension of $Q_p$ to $Q_{p^n} $\cite{Gubser:2016guj,Heydeman:2016ldy}. To avoid clutter we will take $n=1$ although all the expressions can be generalized for generic $n$, basically by replacing $p \to p^n$.  
A $p$-adic number $x$ can be uniquely expressed as an infinite series
\be
x = p^v\sum_{i=0} a_i p^i; \,\, v, a_i \in \mathbb{Z};  \,\,0\le a_i \le p-1; \,\, a_0 \neq 0. 
\ee
The $p$-adic norm is defined as 
\be
|x|_p = p^{-v},
\ee
which satisfies various axioms of norms \cite{norm}.
Conformal symmetry is defined as the transformation 
\be
x \to x'= \frac{ax+b}{cx + d}, \,\, a,b,c,d \in Q_p.
\ee
It furnishes the matrix group PGL$(2,Q_p)$, the direct analogue of SL$(2,\mathbb{R}) $ in 1d conformal transformation in real space-time. 
There are two pieces of algebraic data required to specify completely a $p$-adic CFT \cite{Melzer:1988he}. 
\begin{itemize}
\item First, the spectrum of primary operators $\mathcal{O}_a$ with conformal dimensions $\Delta_a$, 
which transform under conformal symmetry as
\be
\mathcal{O}_a(x) \to \tilde{\mathcal{O}}_a (x') = \bigg\vert\frac{ad-bc}{(cx+d)^2}\bigg\vert^{-\Delta_a}_p  \mathcal{O}_a(x)
\ee. 
\item Second, OPE coefficients $C^{abc}$ defined as \footnote{Note that this expression is exact since $p$-adic CFT is known to have no descendents. }
\be
\mathcal{O}_a(x_1) \mathcal{O}_b(x_2) = \sum_{c} C^{abc}  \vert x_1- x_2\vert_p^{\Delta_c- \Delta_a-\Delta_b} \mathcal{O}_c (x_2).
\ee
The OPE coefficients define an associative operator fusion i.e.$ \sum_c C^{abc} C^{cde} = \sum_c C^{bdc}C^{cae}$. 
There is a unique identity operator $\mathbb{I}$ so that $C^{1ab} = C^{a1b} = \delta_{ab}$,
and that there is a unique dual of $a$ which we denote as $a*$ satisfying $C^{ab1} = \delta_{b a*}$. 
To avoid clutter we work with theories where $a* = a$ and $C^{ab1} = \delta_{ab}$. 
\end{itemize}

\begin{figure}
	\centering
	\includegraphics[width=0.6\linewidth]{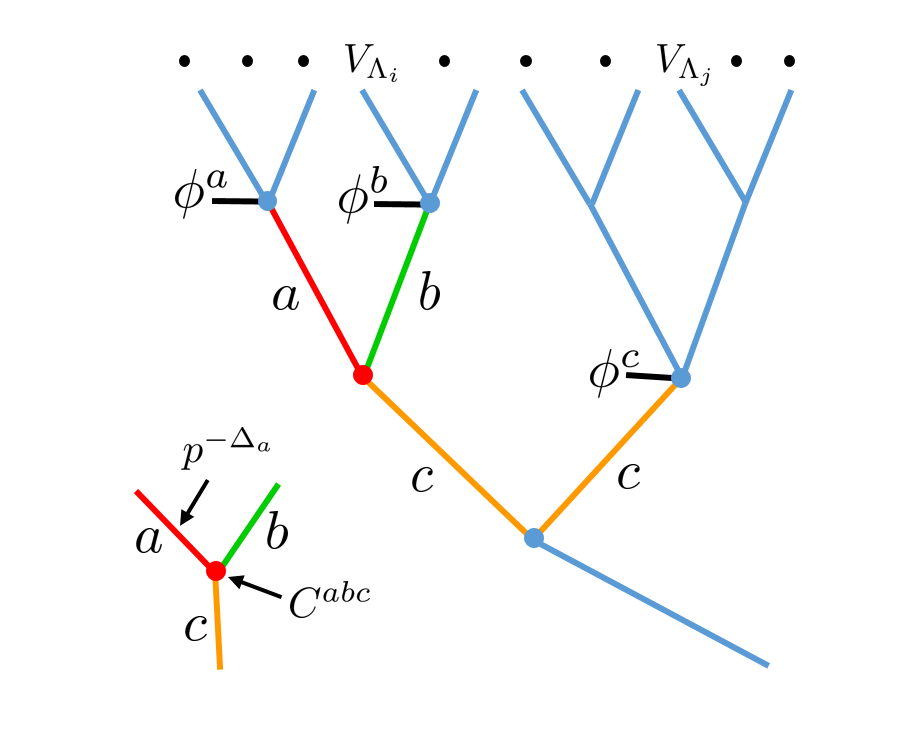}
	\caption{The tensor network representation of a $p=2$-adic CFT. The diagram depicts three bulk operator insertion. Bulk insertions pushed to the asymptotic boundary asymptotes to boundary insertions. The boundary condition $V_{\Lambda_i}$ are chosen to be the fixed point tensor $V^a_f = \delta^a_1$. 
	Each vertex tensor is $C^{abc}$ and each edge of the tensor with index $a$ is weighted by $p^{-\Delta_a}$. }
	\label{fig:ggc}
\end{figure}

{\bf \begin{center} TN on the Bruhat-Tits  tree and the $p$-adic AdS/CFT \end{center}} 
The partition function of a generic $p$-adic CFT can 
be constructed explicitly in the form of a TN covering the Bruhat-Tits (BT) tree \cite{Hung:2019zsk}, 
the latter of which is a discrete $p+1$ valent tree graph. The BT tree is the analogue of the AdS$_{2}$ space whose isometry is the corresponding conformal symmetry group.
 The TN covers the tree, such that at each vertex of the tree sits a $p+1$ index tensor $T^{a_1 \cdots a_{p+1}}$. 
The tensor is given by a fusion tree of $p+1$ operators, expressed in terms of the OPE coefficient of the $p$-adic CFT. 
\be
T^{a_1\cdots a_{p+1}} = \sum_{b_1 \cdots b_{p-2}} C^{a_1 a_2 b_1} C^{b_1 a_3 b_2} \cdots C^{b_{p-2} a_{p} a_{p+1}}.
\ee 
In the special case where $p=2$, $T^{abc} = C^{abc}$.
Two tensors at two vertices connected by an edge are contracted with the edge index weighted by $p^{-\Delta_a}$, where $\Delta_a$ is the conformal dimension of the corresponding primary labeled $a$. 
The asymptotic boundary of the tree is $Q_p$, analogous to of the real line being the asymptotic boundary of the AdS${}_2$ space. 
The partition function is defined by setting appropriate boundary conditions at the cut-off surface which is then taken to approach the asymptotic boundary. Specifically, 
the dangling legs at the cut-off surface $\Lambda$ is projected along the vector $|V_f\rangle \equiv \sum_a \delta^a_0 |a\rangle = |0\rangle  $. 
To compute correlation functions, one inserts operator $\mathcal{O}_a(x_b)$ by projecting the boundary leg at $x_b$ along $|a\rangle$ instead. 
The TN reproduces the correct correlation function in the form of Witten-like diagrams \cite{Hung:2019zsk} . 
The boundary insertions of $\mathcal{O}_a(x_b)$ source these $p^{-\Delta_a}$ weighted paths in the TN, that coincides with  ``bulk-boundary propagators'' $G_a(x_b, v_i)$ on the tree, until
they meet at some vertices $v_i$ in the bulk.  These bulk-boundary propagators are solutions of the graph Klein-Gordon equation
\begin{small}
\begin{eqnarray}
&&(\Box_v +m_a^2) G_a(u,v) = \delta_{u,v}, \, G_a(u,v) = \zeta_p(2\Delta_a)\frac{ p^{-\Delta_a d(u,v)}}{ p^{\Delta_a}}, \\
&&m^2_a =  -\frac{1}{\zeta_p (\Delta_a - n) \zeta_p(-\Delta)}, \,\,\, \zeta_p(s) \equiv \frac{1}{1-p^{-s}} \,\,   \label{eq:mass}  \\
&& \Box_u \phi(u) \equiv \sum_{\langle u v\rangle, {v \sim u}} (\phi(u) - \phi(v)), \,\, \textrm{\tiny $u\sim v \equiv$ nearest neighbour.}
\end{eqnarray}
\end{small}
Here $\langle u v\rangle $ denotes an edge ending at vertices $u,v$, and $d(u,v)$ is the distance between $u,v$.
Here, every link has unit length $d_e =1$.   
Operators labeled $a,b \dots$ traversing different paths can meet at a vertex if they can fuse to the identity
operator. 
The emergence of Witten-diagrams suggests that the TN is recreating the $p$-adic AdS/CFT correspondence proposed in \cite{Gubser:2016guj, Heydeman:2016ldy}, connecting a $p$-adic CFT and a dual bulk theory containing some bulk fields $\phi^a$ in 1-1 correspondence with the primary operators living on the BT tree. 
In fact, we can define bulk field $\phi^a(x)$ insertion by fusing an extra $a$ leg to the bulk vertex $x$. 
The bulk correlation functions $\langle \phi^a(x) \phi^b(y) \cdots \rangle$ are thus
defined as evaluation of the tensor network with the extra legs inserted at the appropriate vertices. See fig \ref{fig:ggc}. 

One can readily show that these results are consistent with an emergent bulk matter field theory living on the BT tree, with action in the large mass limit \footnote{The interaction vertex is essentially fixed at the meeting point of geodesics on the tree rather than summed over \cite{Hung:2019zsk}. Therefore it agrees with the semi-classical limit of a massive field.} 
\begin{small}
\begin{eqnarray} \label{eq:effectiveS}
&S_m &=  \sum_{\langle xy\rangle } \frac{1}{2} \left[  \sum_a (\phi^a(x) - \phi^a(y))^2 +  \frac{m_a^2}{p+1} (\phi^a(x)^2 + \phi^a(y)^2 )\right]   \nonumber \\
&+&  \sum_{a,b,c} \frac{\tilde C^{abc}}{3! (p+1)} [\phi^a(x) \phi^b(x) \phi^c(x) + \phi^a(y) \phi^b(y) \phi^c(y)] +  \cdots, \nonumber  \\
&& \tilde C^{abc}\equiv C^{abc} \sqrt{\frac{p^{\Delta_a + \Delta_b + \Delta_c}}{\zeta_p(2\Delta_a) \zeta_p(2\Delta_b) \zeta_p(2\Delta_c)}},
\end{eqnarray}
\end{small} 
where $\cdots$ corresponds to higher point interaction terms. 
For $p=2$ the action truncates at cubic level.  
We deliberately present the action as a sum over edges anticipating coupling the theory to a dynamical {\it metric} shortly. 

{\bf \begin{center} Distances and Curvatures in a Tensor Network  \end{center}} 
The tensor network constructed appears to be describing a {\it pure BT} space (analogue of pure AdS) when the boundary edges are projected to the identity vector apart from locations where operators are inserted. As in the usual story of AdS/CFT, a most natural way to deform the background geometry is to change the boundary conditions which would drive an RG flow in the CFT. Specifically, each boundary link $i$ at the cutoff surface $\Lambda$ is projected to a generic vector $|V_{\Lambda_i} \rangle$ \cite{Hung:2019zsk}. 
Where there is translation invariance, we can take $|V_{\Lambda_i}\rangle = |V_{\Lambda}\rangle$ for all boundary legs $i$.  Explicitly the vector $|V_{\Lambda}\rangle$ is parametrized as
\be 
|V_\Lambda\rangle = \sum_a V^a_\Lambda |a\rangle. 
\ee
When the vectors $|V_\Lambda\rangle$ are contracted with the  $p$ dangling legs of a tensor $T^{a_1 \cdots a_{p+1}}$ at the cut-off surface, 
it would generate a new vector $|V_{\Lambda -1}\rangle $
\begin{eqnarray}\label{eq:flow1}
&&|V_{\Lambda -1}\rangle \equiv \sum_a V^a_{\Lambda-1} |a\rangle  \nonumber \\
&&= \sum_{a_1\cdots a_{p+1}}V^{a_1}_\Lambda p^{-\Delta_{a_1}}\cdots V^{a_p}_\Lambda p^{-\Delta_{a_p}} T^{a_1 \cdots a_p a_{p+1}} | a_{p+1}\rangle \nonumber \\
\end{eqnarray}
which is fed into the next layer of tensors recursively.  The flow of these vectors  suggests that the geometry described by the tensor network is deformed from the pure BT background.  We note that the original boundary condition describing the un-deformed CFT partition function
\be \label{eq:fixedpt} 
V_f^a = \delta^a_1
\ee
 is indeed a fixed point vector under this flow, which recovers the original pure BT space supposedly dual to the undeformed $p$-adic CFT partition function.  In this case, all the edges contribute equally to the partition function and we can assign unit length to every edge i.e. $d_e = 1$. 

\begin{figure}
	\centering
	\includegraphics[width=0.8\linewidth]{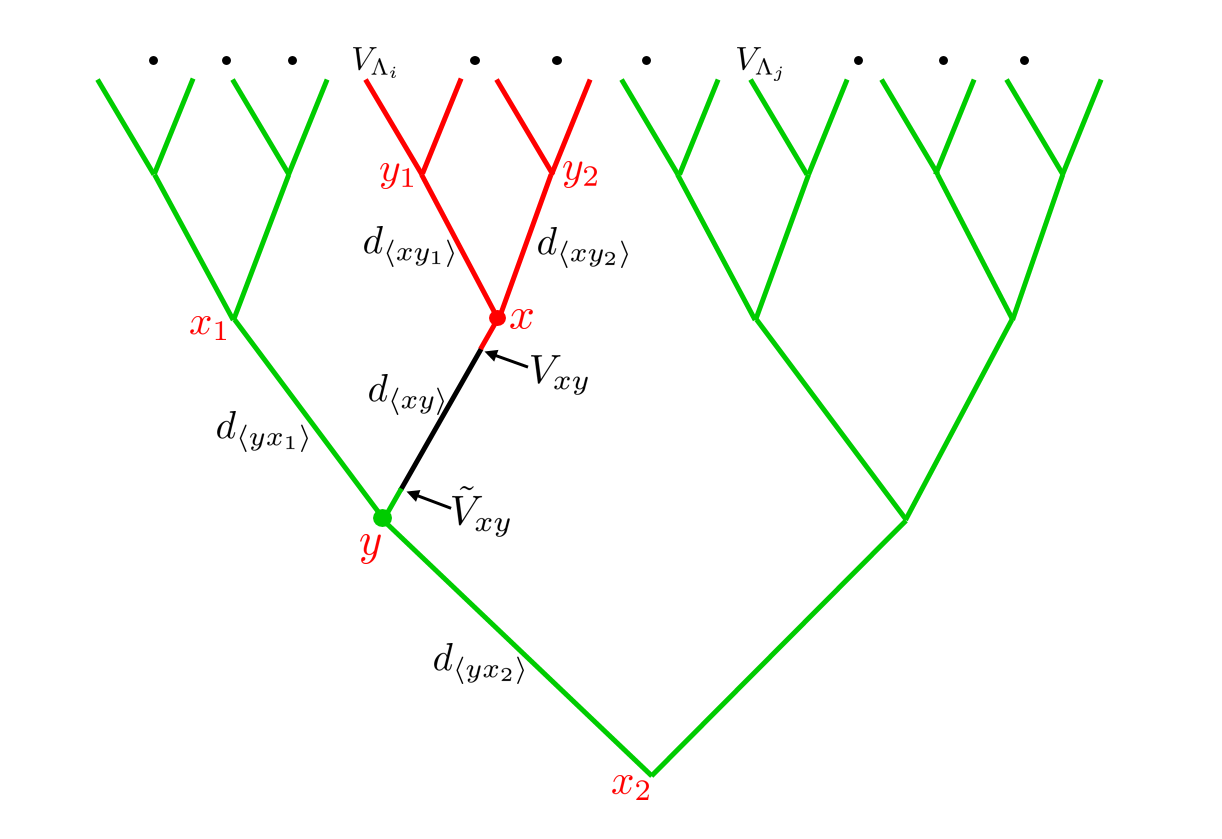}
	\caption{Vector $V^a_{xy}$ and $\tilde V^a_{xy}$ following from the contraction of tensors {\it above}  (colored red) and  {\it below} (colored green) the edge $\langle xy\rangle$ respectively.  Boundary conditions  $V_{\Lambda_i}$  differing from the fixed point vector drives an RG flow. The curvature of the patch centred at $x$ depends on the edge lengths $d_{\langle x y_i\rangle} $ symmetrically. }
	\label{fig:redgreen}
\end{figure}

When we depart from pure BT background, one needs to assign a general length $d_e$ to each edge on the graph. We note that there are two characteristic vectors at an edge $e= \langle xy \rangle$ bounded by vertices $x$ and $y$. One is $|V_{xy} \rangle$ introduced above corresponding to repeating the flow described in (\ref{eq:flow1}), contracting all the tensors from the cutoff surface all the way down to the vertex $x$. 
The other vector $|\tilde V_{xy}\rangle$ follows from analogously contracting all the tensors {\it below} the vertex $y$. This misleading notion of being ``above" or ``below" is illustrated in figure \ref{fig:redgreen}. 

The distinction of $|V_{xy} \rangle$  and $|\tilde V_{xy}\rangle$ comes from boundary conditions very far away,  and so we expect the edge length $d_{\langle xy \rangle}$ to depend on them symmetrically. It is also natural to expect that $|V_{xy} \rangle$  and $|\tilde V_{xy} \rangle$ fully determine $d_{\langle xy \rangle}$. 
To have better analytic control, we consider perturbing around the undeformed CFT by perturbing the boundary conditions $|V_\Lambda\rangle$ around the fixed point tensor: 
\be \label{eq:dV}
V^a_{\Lambda_i} = \delta^a_1 + \lambda v_{\Lambda_i}^a, 
\ee
where $\lambda \ll 1$, but $v^a_{\Lambda_i}$ is completely arbitrary. The boundary conditions are not assumed to respect any translation symmetries.  
In the perturbative limit, the flowed vector in the interior would admit the general expansion
\begin{eqnarray}
V^a_{xy} &=& \delta^a_1 + \omega^a_{xy}, \,\, \omega^a_{xy} \equiv \lambda^{(1)\, a}_{xy} + \lambda^{(2)\, a}_{xy} + \cdots ,  \label{eq:V} \\
{\tilde{V}}^a_{xy} &=&  \delta^a_1 + {\tilde \omega}^a_{xy},\,\,  {\tilde \omega}^a_{xy} \equiv {\tilde \lambda}^{(1)\, a}_{xy} + {\tilde \lambda}^{(2)\, a}_{xy} + \cdots  \label{eq:Vtilde} ,
\end{eqnarray}
where $\lambda^{(n)}, \tilde{\lambda}^{(n)}$ are of order $\lambda^n$. 
The edge length $d_{e= \langle xy \rangle}(V^a_{e}, {\tilde{V}}^a_{e})$ would also admit an expansion about the pure BT space as $d_{e} = 1+ j_e$, where
\begin{eqnarray}\label{eq:j}
  j_e &=& A^a(\omega_e^a+\tilde{\omega}_e^a)+B^{ab}(\omega_e^a\omega_e^b+\tilde{\omega}_e^a\tilde{\omega}_e^b) +C^{ab}\omega_e^a\tilde{\omega}_e^b   \nonumber \\
  &\;& + D^{abc}(\omega_e^a\omega_e^b\omega_e^c+\tilde{\omega}_e^a\tilde{\omega}_e^b\tilde{\omega}_e^c) \nonumber \\
  &&+E^{abc}(\omega_e^a\omega_e^b\tilde{\omega}_e^c+\tilde{\omega}_e^a\tilde{\omega}_e^b\omega_e^c)+\mathcal{O}(\omega^4), 
\end{eqnarray}
for some constants $A^a, B^{ab}, C^{ab} \cdots$. Having assigned edge lengths, one can compute the curvature of this graph. 
There are various proposals. e.g.\cite{Yau1, Ollivier,  Gubser:2016htz}. Consider the curvature $R_x$ of a patch surrounding a vertex $x$. 
The graph curvature should be a symmetric function of lengths $d_{\langle x y_{i= 1, \cdots p+1} \rangle}$ of the $p+1$ edges connected to $x$. 
Generally therefore, in the small $j_e$ limit we expect 
\be \label{eq:gricci}
 R_x = a_0+a_1\sum_i j_{xy_i}+ b\sum_i j^2_{xy_i}+c\sum_{i\neq k}j_{xy_i}j_{xy_k}+\mathcal{O}(j^3),
\ee
again for constants $a_{0,1 }, b,c \cdots $.

{\bf \begin{center} Emergent Action and Einstein equation \end{center}} 
In the above, we have defined distances and curvatures which are, up to some undetermined coefficients,  determined by the TN.
On the other hand, the TN encodes a bulk scalar field theory where the expectation values of the fields $\phi^a(x)$  
can be readily computed. {\bf \it We would therefore like to inquire if the geometry  and the expectation values of $\phi^a(x)$, both read off from the TN,
can be related by some graph Einstein equation. }

To look for such a relation, we need some guidance from an emergent covariant effective bulk action that is consistent with the correlation functions encoded, and obtain the equations of motion following from it, 
before we could even check if such an equation is satisfied by the TN. 
The matter part of the covariant action should reduce to  (\ref{eq:effectiveS}) in the pure BT limit to be consistent with the correlation functions.
 We need to upgrade  (\ref{eq:effectiveS}) by coupling it
to the background geometry via the edge lengths $d_e$. This is attempted in \cite{Gubser:2016htz}, although it has made a choice of treating the mass term  as a local term blind to $d_e$. This is unlike in continuous field theories where every term contains at least the volume form and are thus always sensitive to the metric. 
Rather than making such choices, we write down a more general ansatz  that mimics the continuous covariant action more closely
\begin{small}
\begin{eqnarray}
 S^{cov}_m&=& S^{cov}_2 + S^{cov}_3 + \cdots  \nonumber \\
S^{cov}_2 &=&  \sum_{\langle xy\rangle}d^k_{\langle xy\rangle }({\phi}^a_x-{\phi}^a_y)^2+\sum_{\langle xy\rangle}\frac{d_{\langle xy\rangle}}{p+1}m_a^2(({\phi}^{a}_x)^2+({\phi}^{a}_y)^2) \nonumber \\
S^{cov}_3 &=&\sum_{\langle xy\rangle}\bigg(h(d_{\langle xy\rangle})H^{abc}({\phi}^a_x{\phi}^b_x{\phi}^c_x+{\phi}^a_y{\phi}^b_y{\phi}^c_y) \nonumber \\
&& + r(d_{\langle xy\rangle})R^{abc}({\phi}^a_x{\phi}^b_x{\phi}^c_y+{\phi}^a_y{\phi}^b_y{\phi}^c_x)\bigg),\;\;\;\;
\end{eqnarray}
where $k$ is a constant to be determined. For generality we considered more general cubic interactions other than the exactly local term with coupling $H^{abc}$, and allowed also for nearest neighbour interaction with coupling $R^{abc}$.
The functions $r(d_e)$ and $h(d_e)$ should be regular in the pure BT background. Therefore in the perturbative limit they can be expanded as
\begin{eqnarray}
  h(d_{xy}) &=& h_0+h_1 j_{xy}+h_2 j^2_{xy}+\dots, \\
  r(d_{xy}) &=& r_0+r_1 j_{xy}+r_2 j^2_{xy}+\dots.
\end{eqnarray}
\end{small}
We introduce the {\bf   graph Einstein Hilbert action} making use of the graph curvature introduced in (\ref{eq:gricci}).
\be
S_{EH} = \sum_x R_x(d_{xy_1},d_{xy_2},\dots,d_{xy_{p+1}})+\sum_{\langle xy\rangle} d_{xy} \Lambda,
\ee
where we introduced also the cosmological constant term analogous to $\int d^dx \sqrt{g} \Lambda$. 
The total test effective action is thus $S_{tot} = S_{EH} + S^{cov}_m$. 
Now we are ready to vary these actions wrt each edge length $d_{\langle x y\rangle}$ to obtain the graph Einstein equation. Varying $S_{EH}$ we obtain the {\it graph Einstein tensor } $G$: 
\begin{small}
\begin{eqnarray} \label{eq:G}
 G_{xy}&&\equiv\frac{\delta S_{EH}}{\delta d_{xy}} =  \Lambda+2a_1+4b j_{xy}  \nonumber \\
 &&  +c (\sum_{\substack{i\\(y_i\neq y)}} j_{xy_i} +\sum_{\substack{i\\(x_i\neq x)}}j_{x_iy})+\mathcal{O}(j^2). 
\end{eqnarray}
\end{small}
Varying $S^{cov}_{m}$ we obtain
\begin{small}
\begin{eqnarray} \label{eq:T}
&&  T_{xy}\equiv \frac{\delta S^{cov}_m}{\delta d_{xy}}=
  \frac{k}{2}({\phi}^a_x-{\phi}^a_y)^2+ \frac{m_a^2(({\phi}^{a}_x)^2+({\phi}^{a}_y)^2)}{2(p+1)}
  \nonumber\\
 &\;&+\bigg(h_1H^{abc}({\phi}^a_x {\phi}^b_x {\phi}^c_x+{\phi}^a_y {\phi}^b_y{\phi}^c_y)+
 r_1R^{abc}({\phi}^a_x{\phi}^b_x{\phi}^c_y  \nonumber \\
 && + {\phi}^a_y{\phi}^b_y{\phi}^c_x)\bigg)+  \cdots .
\end{eqnarray}
\end{small}
We would like to substitute the expectation values of $ \phi^a(x) $ into $T$. 
We provide detailed expressions of $ \phi^a(x) $ in the supplementary material. Clearly $ \phi^a(x)  \sim \lambda$, 
and the omitted terms in (\ref{eq:T}) is thus of order $\lambda^4$ and beyond. 
The graph Einstein equation is thus given by 
\be \label{eq:eom}
G_{xy}+T_{xy} =0.
\ee 
We substitute the geometrical data and expectation value of the stress tensor determined by the TN into (\ref{eq:eom}). Requiring that (\ref{eq:eom})
is satisfied order by order in $\lambda$ turns into constraints of the undetermined parameters we have introduced. 
These constraints turn out to be very powerful because $G_{xy}$ and $T_{xy}$ are non-trivial functions of $\lambda^{(n)\, a}_{xy_i}, \tilde \lambda^{(n)\,a}_{xy_i}$ and $\lambda^{(n)\, a}_{yx_i}, \tilde \lambda^{(n)\, a}_{yx_i}$. These edges $\langle x y_i\rangle, \langle y x_i\rangle$ are marked in figure \ref{fig:redgreen} for a given pair of connected vertices $x,y$. These variables $\lambda^{(n)\, a}_{xy_i}, \tilde \lambda^{(n)\,a}_{yx_i}$ are virtually independent because they are distinct functions of generic boundary conditions arbitrarily far away. Therefore, we can isolate the coefficient of each of these independent monomials of $\lambda^{(n)\, a}_{xy_i}, \tilde \lambda^{(n)\,a}_{yx_i}$, and require that it vanishes separately. Up to some overall normalization, it fixes a {\it unique} form of the graph curvature, edge lengths and $S^{cov}_m$ such that (\ref{eq:eom}) can in fact be satisfied. 
Solving constraints up to order $\lambda^2$, we have
\be \label{eq:con1}
\frac{2b}{c} = -p, \,\, A^a = 0, \,\, \Lambda + 2 a_1 = 0.
\ee
Substituting into $G_{xy}$ gives 
$G_{xy} = - c \Box j_{xy} $ which recovers the perturbative graph curvature defined in the Mathematics literature \cite{Yau1,Ollivier,Gubser:2016htz} if we take c= -1. 
Moreover, the matter effective action is constrained to be \begin{eqnarray}
m^2_a &&=  -p-1+p^{1-\Delta_a}+p^{\Delta_a},  \label{eq:mass2}\\
k&&=1.   \label{eq:con3}
\end{eqnarray}
Equation (\ref{eq:mass2}) independently recovers the relation in (\ref{eq:mass}). 
$B^{ab}$ and $C^{ab}$ introduced in (\ref{eq:j}) are in turn determined by $b,c$ and $\Delta_a$. 
Constraints at order $\lambda^3$ again lead to unique expressions for the matter couplings $H^{abc}, R^{abc}$ and 
edge length expansion coefficients $D^{abc}, E^{abc}$ up to an overall undetermined normalization. Detailed expressions are relegated to the appendix. 
We are interested in the limit $m_a^2\to \infty \implies \Delta_a \to \infty $.  This is when the bulk effective action $S_m$ exactly reproduces $\phi^a(x)$ correlation functions of the TN. 
In this limit, satisfyingly only the local term survives with an overall undetermined normalization, 
\be
\lim_{\Delta_a \to \infty, d_e \to 1} S^{cov}_m = -\frac{h_0}{4h_1}\sum_{\langle xy \rangle }  \frac{\tilde C^{abc}}{1+p} \phi^a_x \phi^b_x\phi^c_x,
\ee
but otherwise in exact agreement with the effective action (\ref{eq:effectiveS}).

{\bf \begin{center} Summary and Discussion \end{center}} 
In this paper, we demonstrated that there is, up to some overall normalization,  a {\it unique} way of assigning lengths to the $p$-adic TN so that the geometry read off from the TN satisfies a graph Einstein equation, that is consistent with the bulk effective action that reproduces bulk correlation functions encoded by the TN, in the perturbative limit away from pure BT geometry. We have made minimal assumptions other than locality in our ansatz for the action and also the dependence of edge lengths on the TN data. 
We note that in retrospect, the edge distance $d_{\langle x y\rangle } = 1+ j_{xy}$ defined in (\ref{eq:j}) can be written as
\be
d_{\langle xy \rangle} = 1 - \langle u_x | u_y \rangle,  
\ee
which,  in the limit $\Delta_a \to \infty$
\be
|u_x \rangle =  \frac{1}{\sqrt{2 c(p+1)}}  \sum_{a} {\phi}^a_x|a\rangle. 
\ee
Up to the overall factor in front, this is simply the state corresponding to one dangling leg inserted at a bulk vertex $x$ in the TN. 
Indeed, $d_{\langle xy \rangle}$ ends up being a Fisher information metric between these vertex states.
This is a rare quantitative demonstration of an emergent Einstein equation from a TN that couples to matter, albeit in a simplified setting of $p$-adic CFTs. There are more patterns in the emergent graph Einstein equation that we will report in a forth-coming accompanying paper. We believe this provides further evidence of  geometry being moulded by the correlation of matter, supporting the TN as the microscopic mechanism behind the AdS/CFT correspondence. We believe some ideas and methodology discussed here should admit generalization to TN describing more realistic CFTs. 

\begin{small}
\textit{Acknowledgements.---} LYH acknowledges the support of NSFC (Grant
No. 11922502, 11875111) and the Shanghai Municipal Science and Technology Major Project
(Shanghai Grant No.2019SHZDZX01), and Perimeter Institute for hospitality as a part of the Emmy Noether Fellowship programme. 
Part of this work was instigated in KITP during the program qgravity20.  LC acknowledges support of NSFC (Grant No. 12047515). 
We thank  Bartek Czech, Muxin Han, Ce Shen Gabriel Wong, Qifeng Wu, Mathew Yu and Zhengcheng Gu for useful discussions and comments. 
We thank Si-nong Liu and Jiaqi Lou for collaboration on related projects.
\end{small}

 \bibliography{emergentE}
 
 \newpage
 
 \appendix 

\section{Some details in the perturbative expansion}

Here we would like to supply some extra details of the computations discussed in the main text. 
Further discussion of these results will appear in  a forth-coming company paper. 

The expression of the edge length expanded up to order $\lambda^3$ is given by
\begin{small}
\begin{eqnarray}
  \nonumber
 && d_e
  = 1+B^{ab}({\lambda^{(1)}}_e^a{\lambda^{(1)}}_e^b+\tilde{{\lambda^{(1)}}}_e^a\tilde{{\lambda^{(1)}}}_e^b)
  +C^{ab}{\lambda^{(1)}}_e^a\tilde{{\lambda^{(1)}}}_e^b \nonumber \\
&&  +2B^{ab}({\lambda^{(1)}}_e^a{\lambda^{(2)}}_e^b+\tilde{{\lambda^{(1)}}}_e^a\tilde{{\lambda^{(2)}}}_e^b)
 +C^{ab}({\lambda^{(1)}}_e^a\tilde{{\lambda^{(2)}}}_e^b+{\lambda^{(2)}}_e^a\tilde{{\lambda^{(1)}}}_e^b)   \nonumber \\
&&  +D^{abc}({\lambda^{(1)}}_e^a{\lambda^{(1)}}_e^b{\lambda^{(1)}}_e^c  +\tilde{{\lambda^{(1)}}}_e^a\tilde{{\lambda^{(1)}}}_e^b\tilde{{\lambda^{(1)}}}_e^c) \nonumber \\
&& +E^{abc}({\lambda^{(1)}}_e^a{\lambda^{(1)}}_e^b\tilde{{\lambda^{(1)}}}_e^c+\tilde{{\lambda^{(1)}}}_e^a\tilde{{\lambda^{(1)}}}_e^b{\lambda^{(1)}}_e^c)+\mathcal{O}(\lambda^4).
\end{eqnarray}
\end{small}

The expectation value of the bulk field  $ \tilde \phi^a_x \equiv \sqrt{\frac{p^{\Delta_a}}{\zeta_p(2\Delta_a)}} \phi^a_x $
at $x$ and a neighbour $y$
can be expressed in terms of the characteristic vectors $V_{i}, \tilde V_{i}$ (see figure \ref{part}) which 
can be expanded in $\lambda$ as follows
\begin{small}
\begin{eqnarray}
\label{phixa1}
&&  \tilde{\phi}_x^a ={\lambda^{(1)}}^a p^{-\Delta_a}+\tilde{{\lambda^{(1)}}}^a p^{-2\Delta_a}+{\lambda^{(2)}}^a p^{-\Delta_a}+\tilde{{\lambda^{(2)}}}^a p^{-2\Delta_a} \nonumber \\
 && +\gamma^a+\tilde{\gamma}^a p^{-\Delta_a}+{\lambda^{(1)}}^b\tilde{{\lambda^{(1)}}}^c p^{-\Delta_b}p^{-2\Delta_c}C^{abc}+\mathcal{O}({\lambda}^3),\;\;\\
  \label{phiya1}
&&  \tilde{\phi}_y^a =\tilde{{\lambda^{(1)}}}^a p^{-\Delta_a}+{\lambda^{(1)}}^a p^{-2\Delta_a}+\tilde{{\lambda^{(2)}}}^a p^{-\Delta_a}+{\lambda^{(2)}}^a p^{-2\Delta_a} \nonumber \\
&& +\tilde{\gamma}^a+\gamma^a p^{-\Delta_a}+\tilde{{\lambda^{(1)}}}^b{\lambda^{(1)}}^c p^{-\Delta_b}p^{-2\Delta_c}C^{abc}+\mathcal{O}({\lambda}^3),\;\;
\end{eqnarray}
\end{small}
where
\begin{eqnarray}
  {\lambda^{(1)}}^a &\equiv& \sum_{i=1}^p {\lambda^{(1)}}^a_i,\\
  \tilde{{\lambda^{(1)}}}^a &\equiv& \sum_{i=1}^p \tilde{{\lambda^{(1)}}}^a_i. \\
  {\lambda^{(2)}}^a &\equiv& \sum_{i} {\lambda^{(2)}}^a_i,\\
  \tilde{{\lambda^{(2)}}}^a &\equiv& \sum_{i} \tilde{{\lambda^{(2)}}}^a_i,\\
  \gamma^a&\equiv& \sum_{i\neq j, b,c}{\lambda^{(1)}}^b_i {\lambda^{(1)}}^c_j C^{abc}p^{-\Delta_b}p^{-\Delta_c},\\
  \tilde{\gamma}^a&\equiv& \sum_{i\neq j, b,c}\tilde{{\lambda^{(1)}}}^b_i \tilde{{\lambda^{(1)}}}^c_j C^{abc}p^{-\Delta_b}p^{-\Delta_c}.
\end{eqnarray}

We remind the readers that 
\begin{eqnarray}
V^a_{i} &=& \delta^a_1 + \omega^a_{i}, \,\, \omega^a_{i} \equiv \lambda^{(1)\, a}_{i} + \lambda^{(2)\, a}_{i} + \cdots ,   \\
{\tilde{V}}^a_{i} &=&  \delta^a_1 + {\tilde \omega}^a_{i},\,\,  {\tilde \omega}^a_{i} \equiv {\tilde \lambda}^{(1)\, a}_{i} + {\tilde \lambda}^{(2)\, a}_{i} + \cdots   ,
\end{eqnarray}
See figure \ref{part} for the characteristic vector each $V_i$ corresponds to .  

\begin{figure}[htbp!]
\centering

\label{part}
\includegraphics[width=0.3\textwidth]{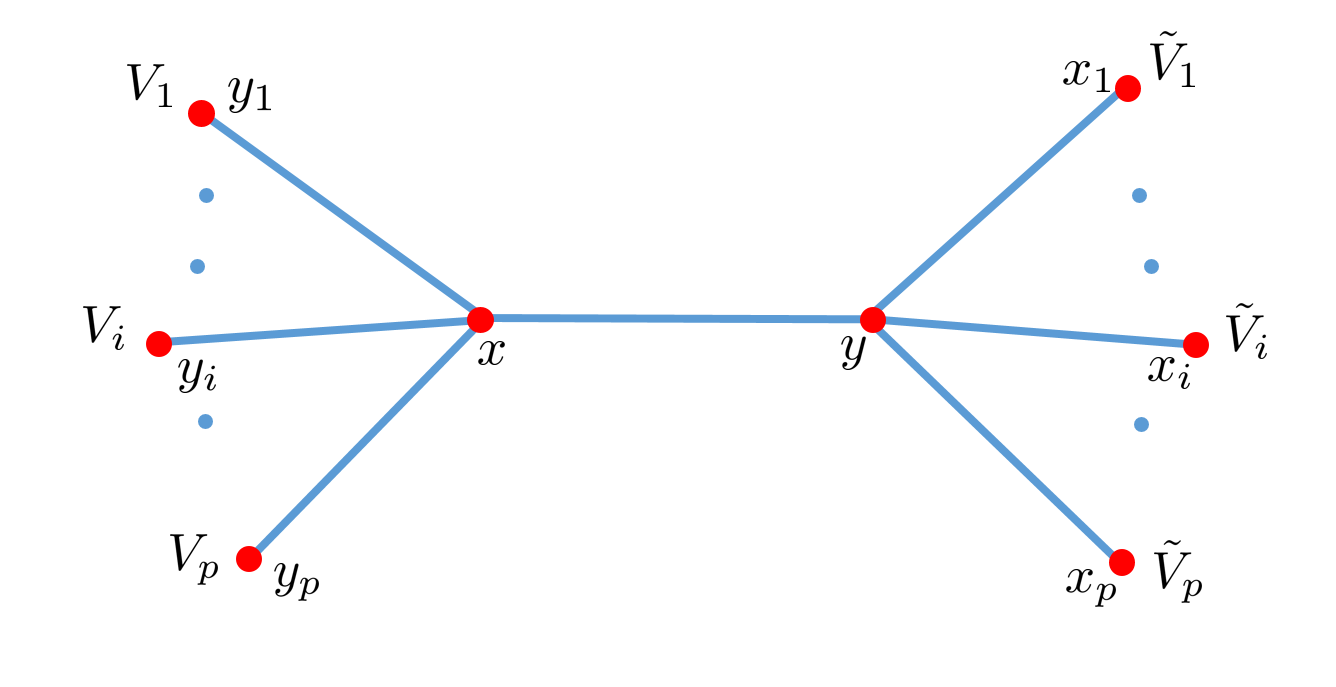}

\caption{The edge $xy$ is connected with $\{xy_i\}$ and $\{x_iy\}$. When one restricts attention to this patch, $\{V_i^a\}$ and $\{\tilde{V}_i^a\}$ encodes all needed information.}
\label{part}

\end{figure}

The {\it graph Einstein tensor} $\delta S_{EH}/\delta d_{\langle xy \rangle }$ up to order $\lambda^2$ is given by
\begin{eqnarray}
 &&G_{xy} =
 \sum_{a,b} p^{-2 (\Delta_a+\Delta_b)} ({\lambda^{(1)}}^a{\lambda^{(1)}}^b+\tilde{{\lambda^{(1)}}}^a\tilde{{\lambda^{(1)}}}^b) \bigg(4 b B^{ab} p^{\Delta_a+\Delta_b} \nonumber \\
&&  +B^{ab} c \left(-2 p^{\Delta_a+\Delta_b}+p^{\Delta_a+\Delta_b+1}+p\right) +c C^{ab} p^{2 \Delta_a+\Delta_b}\bigg)   \nonumber \\
&&+p^{-2 (\Delta_a+\Delta_b)} {\lambda^{(1)}}^a\tilde{{\lambda^{(1)}}}^b  \nonumber \\
&& \bigg(4 b C^{ab} p^{\Delta_a+\Delta_b}+2 B^{ab} c (p-1) \big(p^{\Delta_b}+p^{\Delta_a}\big) \nonumber \\
&&+c C^{ab} \big(p^{2 \Delta_a}+p^{2 \Delta_b}\big)\bigg) +\sum_i \frac{c({\lambda^{(1)}}^a_i{\lambda^{(1)}}^b_i+\tilde{{\lambda^{(1)}}}^a_i\tilde{{\lambda^{(1)}}}^b_i)}{2}  \nonumber \\
&&\bigg(2B^{ab}(1+p^{-\Delta_a-\Delta_b})
-C^{ab} (p^{-\Delta_a}+p^{-\Delta_b})\bigg)  \nonumber \\
&&+\mathcal{O}({\lambda}^3).
  \end{eqnarray}

The stress tensor defined as $\delta S^{cov}_m /\delta d_{xy}$ up to order $\lambda^2$ is given by

 \begin{eqnarray}
&&  T_{xy} =\sum_a p^{-3 \Delta_a} ({\lambda^{(1)}}^a{\lambda^{(1)}}^a+\tilde{{\lambda^{(1)}}}^a\tilde{{\lambda^{(1)}}}^a) \nonumber \\
&&\frac{\bigg(k (p+1) \bigg(p^{\Delta_a}-1\bigg)^2+m_a^2 \big(p^{2 \Delta_a}+1\big)\bigg)}{2(p+1) \big(p^{\Delta_a}-1\big) \big(p^{\Delta_a}+\big)} \nonumber \\
  &\;&+ \frac{p^{-3 \Delta_a} \bigg(2 m_a^2 p^{\Delta_a}-k (p+1) \big(p^{\Delta_a}-1\big)^2\bigg)}{(p+1) \big(p^{\Delta_a}-1\big) \big(p^{\Delta_a}+1\big)}{\lambda^{(1)}}^a\tilde{{\lambda^{(1)}}}^a \nonumber \\
  &&+\mathcal{O}(\lambda^3).
\end{eqnarray}

Let us give an example how the equations of motion leads to powerful constraints on the couplings. 
By requiring that $G_{xy} + T_{xy} = 0$, one can see that the monomial $\lambda_i^{(1)\, a} \lambda_i^{(1)\, a}$ only appears in $G$ but not in $T$. Therefore its
coefficient must be vanishing by itself. 
\be
 \frac{c}{2} \bigg(2B^{ab}(1+p^{-\Delta_a-\Delta_b})-C^{ab} (p^{-\Delta_a} 
  +p^{-\Delta_b})\bigg) = 0,
\ee

We subsequently consider the coefficients of $\lambda^{(1)\, a} \lambda^{(1)\, a} $, ${\tilde \lambda}^{(1)\, a} {\tilde \lambda}^{(1)\, a}$
and ${\lambda}^{(1)\, a} {\tilde \lambda}^{(1)\, a}$ separately to obtain the constraints (\ref{eq:con1}, \ref{eq:mass2}, \ref{eq:con3}) in the main text. 
In addition we get
\begin{eqnarray}
  B^{aa} = \frac{1}{2 c (p+1) \left(1-p^{2 \Delta_a}\right)}.
\end{eqnarray}

Then we expand $G$ and $T$ up to order $\lambda^3$ to solve for constraints
at that order. We will not reproduce all the expressions here, but present the solutions of the constraints. 
 
 \begin{eqnarray}
\label{Rabc1}
&&  R^{abc} = \frac{\tilde C^{abc}\left(p^{\Delta _a+\Delta _c}-p^{\Delta _b}\right) \left(p^{\Delta _b+\Delta _c}-p^{\Delta _a}\right)}{2 r_1 \left(p^{2 \Delta _a}-1\right) \left(p^{2 \Delta _b}-1\right) \left(p^{2 \Delta _c}-1\right)}, \nonumber \\
  \label{Eabc}
&&  E^{abc}=\frac{\tilde C^{abc}}{4 c (p+1) \left(p^{2 \Delta _a}-1\right) \left(p^{2 \Delta _b}-1\right) \left(p^{2 \Delta _c}-1\right)} \nonumber \\
&& ( -p^{\Delta _a+\Delta _b+2 \Delta _c}-p^{\Delta _a-\Delta _b}-p^{\Delta _b-\Delta _a}+3 p^{\Delta _a+\Delta _b}\nonumber \\
&&-p^{2 \Delta _a+\Delta _c}-p^{2 \Delta _b+\Delta _c}-p^{-\Delta _c}+3 p^{\Delta _c}).
\end{eqnarray}

Having $E^{abc},R^{abc}$, we can further get
\begin{small}
\begin{eqnarray}
\nonumber
 D^{abc}&=&\frac{- \tilde C^{abc}p^{-\Delta _a-\Delta _b-\Delta _c}}{12 c (p+1) \left(p^{2 \Delta _a}-1\right) 
   \left(p^{2 \Delta _b}-1\right) \left(p^{2 \Delta _c}-1\right)} \nonumber \\
&&   \Big(-3 p^{\Delta _a+\Delta _b+\Delta _c}-3 p^{2 \left(\Delta _a+\Delta _b+\Delta _c\right)}   \nonumber \\
  &\;&+p^{3 \Delta _a+\Delta _b+\Delta _c}+p^{\Delta _a+3 \Delta _b+\Delta _c}+p^{\Delta _a+\Delta _b+3 \Delta _c} \nonumber \\
  &&+p^{2 \left(\Delta _a+\Delta _b\right)}+p^{2 \left(\Delta _a+\Delta _c\right)}    +p^{2 \left(\Delta _b+\Delta _c\right)}\Big),
  \label{Dabc1}\\
  \nonumber
H^{abc}&=&\frac{\tilde C^{abc}p^{-\Delta _a-\Delta _b-\Delta _c}  \bigg(p^{\Delta _a+\Delta _b+\Delta _c}+p\bigg)}{12 h_1 (p+1) \big(p^{2 \Delta _a}-1\big) \big(p^{2 \Delta _b}-1\big) \big(p^{2 \Delta _c}-1\big)}  \nonumber \\
&&  \Big(-3 p^{\Delta _a+\Delta _b+\Delta _c}-3 p^{2 \big(\Delta _a+\Delta _b+\Delta _c\big)}   +p^{3 \Delta _a+\Delta _b+\Delta _c} \nonumber \\
&&+p^{\Delta _a+3 \Delta _b+\Delta _c}+p^{\Delta _a+\Delta _b+3 \Delta _c}+p^{2 \left(\Delta _a+\Delta _b\right)} \nonumber \\
&&+p^{2 \left(\Delta _a+\Delta _c\right)}+p^{2 \left(\Delta _b+\Delta _c\right)}\Big).
  \label{Habc1}
\end{eqnarray}
\end{small}

A very simple relation between these coupling constants is found:
\begin{eqnarray}
  \frac{D^{abc}}{ H^{abc}} &=& \frac{-h_1}{c(p^{\Delta_a+\Delta_b+\Delta_c}+p)}.
\end{eqnarray}

Finally, as already noted in the main text, on retrospect  $d_{\langle u v \rangle}$ can be written as  $1 - \langle u| v\rangle$
for some emergent states $|u\rangle, |v\rangle$. Before taking the {\it semi-classical} limit $\Delta_a \to \infty$, the complete expressions for them are given by
\begin{eqnarray}
  |u\rangle &\equiv& \sum_a (\tilde{\phi}^a_u+\tilde{\phi}^b_u\tilde{\phi}^c_u \tilde{R}^{bca} )|\tilde a\rangle,\\
  |v\rangle &\equiv& \sum_a (\tilde{\phi}^a_v+\tilde{\phi}^b_v\tilde{\phi}^c_v \tilde{R}^{bca} )| \tilde a\rangle,
\end{eqnarray}
where
\begin{eqnarray}
  \langle \tilde a | \tilde b\rangle &=& \frac{-\delta^{ab}p^{\Delta _a}}{2 c (p+1) \left(1-p^{2 \Delta _a}\right)},\\
  \tilde{R}^{abc}&\equiv&{r_1R^{abc}}  \sqrt{\frac{\zeta_p(2\Delta_a) \zeta_p(2\Delta_b)}{ \zeta_p(2\Delta_c) p^{\Delta_a + \Delta_b - \Delta_c} } }.
\end{eqnarray}

\end{document}